**Comment on "Nematic Electronic Structure in the "Parent" State of the Iron-Based Superconductor Ca(Fe$_{1-x}$Co$_x$)$_2$As$_2$"**

Simon A.J. Kimber[1*], Dimitri N. Argyriou[1*] and I.I. Mazin[2+]

[1] Helmholtz-Zentrum Berlin für Materialien und Energie, Hahn-Meitner Platz 1, Berlin 14109, Germany
[2] Navel Research Laboratory, code 6390, 4555 Overlook Avenue Southwest, Washington DC 20375, USA.

*simon.kimber@helmholtz-berlin.de, *argyriou@helmholtz-berlin.de, +mazin@nrl.navy.mil

**Chuang et al (Reports, 8th of January 2010, p. 181)[1] report quasiparticle interference (QPI) imaging that shows pronounced $C_2$ asymmetry. They interpreted this result as indication of an electronic nematic phase with an electron band dispersive only along the *b*-axis of this orthorhombic material. We argue that this asymmetry is consistent with the underlying long range magnetic order and that LDA electronic structure provides a better description of the QPI images than the 1D band structure conjectured by Chuang et al.**

Clarifying the origin of magnetism in the Fe based superconductors is of significant importance to the condensed matter community as it will set the correct physical framework upon which to construct model for their high-Tc superconductivity. While superconductivity in the high-Tc cuprates arises by doping a Mott insulator, the parent phases of the Fe based superconductors are metallic with magnetic order consistent with the nesting of the Fermi surface evident in the LDA calculations[2] and confirmed directly by ARPES measurements[3].

Nematic phases are frequently found in organic matter. The defining characteristic of these phases is orientational order in the absence of long range positional order, resulting in distinctive uniaxial physical properties. It has also been proposed that nematic order exists in some electronic systems, and may even play a role in mediating high temperature superconductivity[4]. Borzi et al demonstrated the presence of just such a phase in Sr$_3$Ru$_2$O$_7$ at millikelvin temperatures and high magnetic fields[5]. Importantly, this compound was shown to remain strictly tetragonal with $C_4$ structural symmetry, while a pronounced $C_2$ asymmetry in electronic properties was measured. This breaking of the electronic symmetry compared to that of the underlying lattice is a hallmark of electronic nematic phases.

Such a hallmark is not apparent in the QPI imaging reported by Chuang et al.[1], on a lightly cobalt doped sample of CaFe$_2$As$_2$, which is a parent compound of the newly discovered iron arsenide superconductors. The reported STM examination shows a momentum space electronic structure with completely broken $C_4$ symmetry and dispersive electronic excitations along the *b* axis and a q=2π/8d$_{Fe-Fe}$ band folding along the *a* axis. This is taken to mean " that the scattering interference modulations are strongly unidirectional, which should occur if the *k*-space band supporting them is nematic". However these features were observed in an orthorhombic lattice where $C_2$ symmetry is already seriously broken. Indeed the size of the orthorhombic distortion is not "minute" with *b*/*a* ~ 1 % and is instead comparable with distortions seen in various iron oxides systems. For instance, at the Verwey transition the Fe-O bond dilation is ~0.6% with Fe atoms in the same tetrahedral symmetry as in the ferropnictide superconductors[6], and this is usually considered to be a strong distortion. Similarly, in the antiferromagnetic phase of FeO, where the cubic symmetry is completely broken, the structural effect is also of the same order[7].

Since the tetragonal symmetry is decisively broken at the onset of the magnetic order in this ferropnictide, it is completely clear that the symmetry of the electronic structure defining the structural distortion is also broken. While it is argued that the a "Pomeranchuk instability toward a $C_2$ symmetry nematic electronic structure is possible", the cited reference[8] discusses the changes in band structure at the tetragonal-orthorhombic structural phase transition, where the Fermi surface loses four-fold rotational symmetry. Moreover, while the observed QPI pattern does violate the $C_4$ symmetry, it is clearly not 1D, in the sense that it varies equally strongly along $k_x$ and $k_y$ directions. Thus, interpretation of the data in terms of a 1D electron band does not appear to be possible. On the other hand, it appears that they have a clear and more conventional explanation in terms of the calculated LDA electronic structure.

Indeed, antiferromagnetic ordering has a strong effect on the Fermi surface of Fe pnictides (Fig. 1a), as manifested by the first principle calculations[9]. Apart from small quasi-2D tubular pockets, originating from Dirac cones, there is one hole pocket around Z (0,0,_/c or 2_/a,0,0) and two electron pockets between Z and 0, _ /b, _ /c. Electron-hole quasiparticle scattering is particularly effective and gives rise[9] to a characteristic pattern (Fig. 1b) very similar to the one obtained in Ref. (1), with two sharp maxima at q=0,±ξ,0, where ξ~ _ /4b. Note that the authors of Ref. (1) were not able to detect any Ca ions on the surface, so the sample surface is likely charged with up to 0.5 hole per Fe. For this reason, and due to the absence of many-body renormalization in the calculations, it does not make sense to pursue a quantitative agreement between the present experiment and these calculations. Yet, these LDA calculations without any adjustable parameters are in excellent qualitative agreement with the QPI images. Note that while the calculated Fermi surfaces completely break the tetragonal symmetry, the individual pockets are very 3D, so that the calculated conductivity is comparable for all three directions. While experimentally there is up to a 20% *a/b* charge transport anisotropy close to the tetragonal to orthorhombic phase boundary in $CaFe_2As_2$ [9,10], it is much less than what would be predicted for a quasi 1D electronic band, and of the opposite sign.

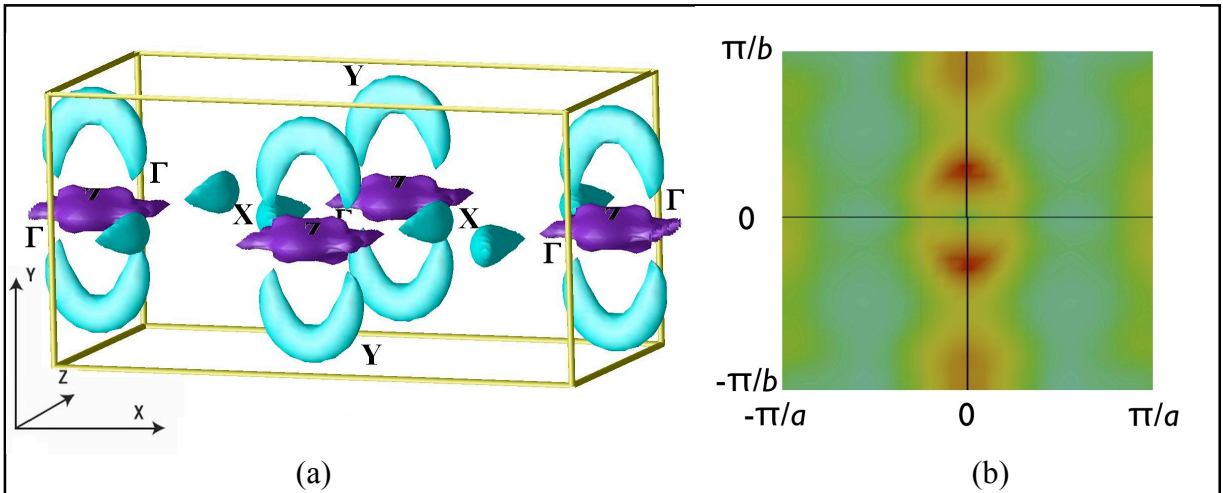

(a)          (b)

Fig. 1. (a) Calculated Fermi surface of $CaFe_2As_2$ in the antiferromagnetic state. (b) Quasiparticle interference pattern for zero bias and $q_z$~0, calculated using the same electronic structure and Equation S5 of reference 9.

To summarize, we have shown that (a) the QPI interference reported by Chuang et al. can be explained as a result of the band structure change brought forth by the long range antiferromagnetic ordering and is unrelated to a possible nematic state above the magnetic transition, and (b) while the resulting band structure manifests strongly an orthorhombic symmetry, it is not at all quasi-1D.